\documentclass[preprint,showpacs,aps,preprintnumbers,amsmath,amssymb,
nofootinbib]{revtex4}
\usepackage{axodraw}
\usepackage{epsfig}
\begin{document}

\begin{flushright}
\end{flushright}


\newcommand{\be}{\begin{equation}}
\newcommand{\ee}{\end{equation}}
\newcommand{\bea}{\begin{eqnarray}}
\newcommand{\eea}{\end{eqnarray}}
\newcommand{\bers}{\begin{eqnarray*}}
\newcommand{\eers}{\end{eqnarray*}}
\newcommand{\nn}{\nonumber}
\newcommand\un{\cal{U}}
\def\Apa{A_\parallel}
\def\Ape{A_{\perp}}
\def\lp{\lambda^\prime}
\def\ll{\Lambda}
\def\mb{m_{\Lambda_b}}
\def\ml{m_\Lambda}
\def\s1{\hat s}
\def\ds{\displaystyle}
\def\s{\smallskip}
\def\l{\hspace*{0.05cm}}
\def\esp{\hspace*{1cm}}

\title{\large $B_s^0 - \overline{B}_s^0$
mixing and  $b \to s $ transitions in isosinglet down quark model}
\author{R. Mohanta$^1$  and A. K. Giri$^2$  }
\affiliation{$^1$ School of Physics, University of Hyderabad,
Hyderabad - 500 046, India\\
$^2$ Department of Physics, Punjabi University,
Patiala - 147 002, India}

\begin{abstract}
The recent observation of the mass difference in $B_s$ system seems
to be not in complete agreement  with the corresponding standard
model value. We consider the model with an extra vector like down
quark to explain this discrepancy and obtain the constraints on the
new physics parameters. Thereafter, we show that with these new
constraints this model can successfully explain other observed
deviations associated with $b \to s$ transitions, namely, $B_s \to
\psi \phi$, $B\to K \pi$ and $B\to \phi K_s$.
\end{abstract}

 \maketitle

\section{Introduction}

The results of the currently running two asymmetric $B$ factories
confirmed the fact that the phenomenon of CP violation in the
Standard Model (SM) is due to the complex phase in the CKM quark
mixing matrix \cite{ckm}. The observed data are almost in the line
of the SM expectations and there is no clear indication of new
physics so far. However, there are some interesting deviations from
that of the SM expectations which could provide us an indirect
signal of new physics. Here we are concentrating on few such
deviations which are associated with the CP violation parameters of
flavor changing neutral current (FCNC) mediated $b \to s $
transitions.
 A partial list includes:

$\bullet$ The observed mass difference measured between heavy and
light $B_s$ mesons \cite{cdf1} seems to be inconsistent with its SM
value with a deviation of few sigma.

$\bullet $ The observed discrepancy between the measured $S_{\phi
K_s}$ and $S_{\psi K_s}$ \cite{hfag} already gave an indication of
the possible existence of NP in the $B \to \phi K_s$ decay
amplitude. Within the SM, these CP symmetries are expected to be
same with a deviation of about $5 \%$ \cite{yg1}.

 $\bullet$ The
recent observation of a very large $S_{\psi \phi}$ by the CDF
collaboration \cite{cdf2} is in contrast to its expected SM value
i.e., $S_{\psi \phi} \approx 0$. This may be considered as a clear
signal of new physics in the $b \to s$ transitions.

$\bullet $  There appears to be some disagreement between the direct
CP asymmetry parameters of $B^-\to \pi^0K^- $ and that of the $ \bar
B^0\to \pi^+K^-$ . $\Delta A_{CP}(K \pi)$, which is the difference of
these two parameters, is found to be around 15\% \cite{hfag}, whereas the SM
expectation is vanishingly small. This constitutes what is called 
$\pi K$ puzzle in the literature and is believed to be an indication of
the existence of new physics.

 $\bullet $  $B_s\to \mu^+ \mu^-$ problem has been widely discussed
in the literature. The SM value is quite small (we have only upper
limit for the branching ratio) and it is very clean mode so if we
have any smoking gun signal of new physics elsewhere in $ b\to s$
transitions it is quite likely that it could also be found in this
mode. Therefore, $B_s\to \mu^+ \mu^- $ is a golden mode to detect new
physics.

In this paper, we would like to see the effect of the extra vector
like down quark \cite{yg2} in explaining the above mentioned
observed discrepancies. It is a simple model beyond the standard
model with an enlarged matter sector due to an additional vector
like down quark $D_4$.
 Isosinglet
quarks appear in many extensions of the SM like the low energy limit
of the $E_6$ GUT models \cite{e6}. The mixing of this singlet type down quark
with the three SM down type quarks  provides a framework to study
the deviations of the unitarity constraint of the $3 \times 3 $ CKM
matrix. To be more explicit, the presence of an additional
down quark implies a $4 \times 4$ matrix $V_{i \alpha}$ $(i=u,c,t,4,
 ~ \alpha=d,s,b, b^\prime)$ would diagonalize the down quark
mass matrix. Due to this, some new features appear in the low energy
phenomenology. The charged currents are unchanged except that the
$V_{CKM}$ is now the $3 \times 4$ upper sub-matrix of $V$. However,
the distinctive feature of this model is that the FCNC interaction
enters at tree level in the neutral current Lagrangian of the left
handed down quarks as \cite{yg2} \be {\cal L}_Z= \frac{g}{2 \cos
\theta_W}\Big[\bar u_{Li} \gamma^\mu u_{Li}-\bar d_{L
\alpha}U_{\alpha \beta} \gamma^\mu d_{L \beta}-2 \sin^2 \theta_W
J^\mu_{em}\Big] Z_\mu\;, \label{fcnc}\ee with \be U_{\alpha \beta}=
\sum_{i=u,c,t} V_{\alpha i}^\dagger V_{i \beta}= \delta_{\alpha
\beta}-V_{4 \alpha}^* V_{4 \beta}\;, \ee where $U$ is the neutral
current mixing matrix for the down sector, which is given above. As
$V$ is not unitary, $U \neq {\bf 1}$. In particular the non-diagonal
elements do not vanish \be U_{\alpha \beta}=-V_{4 \alpha}^* V_{4
\beta} \neq 0 ~~~~ {\rm for}~~~\alpha \neq \beta\;. \ee Since the
various $U_{\alpha \beta}$ are non vanishing they would signal new
physics and the presence of FCNC at the tree level which can
substantially modify the predictions of SM for the FCNC processes.
Of course, these low energy couplings are severely restricted by the
low energy results available on  different FCNC processes
i.e., $Br(K_L \to \mu \bar \mu)_{\rm SD}$, $Br(K^+ 
\to \pi^+ \nu \bar \nu)$, $\epsilon_K$, $\Delta M_K$,
$\Delta M_{B_d}$, $\Delta M_{B_s}$, $Br(B \to X_{d,s} ~l^+
l^-$ etc \cite{bbm}.
Nevertheless, it is well known that even fulfilling these strong
constraints there could still be large effects on $B$ factory
experiments on CP violation. The implications of the FCNC mediated
$Z$ boson effect has been extensively studied in the context of $b$
physics \cite{rm, desh, vives}.

\section{$B_s -\bar B_s$ Mixing}

We will first concentrate  on the mass difference between the
neutral $B_s$ meson mass eigenstates ($\Delta M_s$) that
characterizes the $B_s -\overline{B}_s$ mixing phenomena. In the SM,
$B_s - \overline{ B}_s$ mixing occurs at the one-loop level  by
flavor-changing weak interaction box diagrams and hence is very
sensitive to new physics effects.

In the SM, the effective Hamiltonian describing the $\Delta B=2$
transition,  induced by the box diagram, is given by \cite{lim} \be {\cal
H}_{eff}=\frac{G_F^2}{16 \pi^2}~ \lambda_t^2~ M_W^2 S_0(x_t)\eta_t
(\bar s b)_{V-A}(\bar s b)_{V-A} \ee where $\lambda_t=V_{tb}
V_{ts}^*$, $\eta_t$ is the QCD correction factor and $S_0(x_t)$ is
the loop function \be S_0(x_t)=\frac{4 x_t -11 x_t^2
+x_t^3}{4(1-x_t)^2} - \frac{3}{2} \frac{\log x_t x_t^3}{(1-x_t)^3},
\ee with $x_t=m_t^2/M_W^2$. Thus, the $B_s - \bar B_s$ mixing
amplitude in the SM can be written as \be M_{12}^{SM}=\frac{1}{2
M_{B_s}} \langle \bar B_s|{\cal H}_{eff}| B_s \rangle = \frac{G_F^2}{12
\pi^2} M_W^2~ \lambda_t^2~ \eta_t~ B_s f_{B_s}^2 M_{B_s} S_0(x_t)\;,
\label{sm} \ee where the vacuum insertion method has been used to
evaluate the matrix element \be\langle \bar B_s|(\bar s
b)_{V-A}(\bar s b)_{V-A}|B_s \rangle = \frac{8}{3} B_s f_{B_s}^2
M_{B_s}^2\;.\label{vac}\ee The corresponding mass difference is
related to the mixing amplitude through $\Delta M_s = 2 |M_{12}|$.

Recently, Lenz and Nierste \cite{lenz} updated the theoretical
estimation of the $B_s$ mass difference in the SM, with the value
$(\Delta M_{B_s})^{\rm SM}= (19.30 \pm 6.68)~ {\rm ps}^{-1}$ (for
Set-I parameters) and $(\Delta M_{B_s})^{\rm SM}= (20.31 \pm 3.25)~
{\rm ps}^{-1}$ (Set-II).

The CDF \cite{cdf1} and D\O~ \cite{d0} collaborations have also
recently  reported new results for the $B_s - \bar{B}_s $ mass
difference \bea 
&& \Delta M_{B_s}=(17.77 \pm 0.10 \pm 0.07)~ {\rm
ps^{-1}}~~~~~~~~~({\rm CDF})\nn\\
&& 17 ~{\rm ps^{-1}} < \Delta M_{B_s} < 21~ {\rm
ps^{-1}}~~~~~~~~90 \%~ {\rm C.L.}~
({\rm D \O })\;.
 \eea

Although the experimental results appear to be consistent with the
standard model prediction, but they do not completely exclude the
possible new physics effects in $\Delta B=2$ transitions. In the
literature, there have already been many discussions both in model
independent \cite{ball1, np, np1} and model dependent way
\cite{susy} regarding the implications of these new measurements. In
this work we would like to see the effect of the extended isosiglet
down quark  model on the mass difference of $B_s$ system and its
possible implications for the other $b \to s$ transition processes.


\begin{figure}[t]
\begin{center}
\begin{picture}(300,50)(0,0)
\SetColor{Black}
\Photon(50,25)(100,25){5}{4}
\SetColor{Black}
\ArrowLine(20,0)(50,25){}{}
\SetColor{Black}
\ArrowLine(50,25)(20,55){}{}\Vertex(50,25){2}
\SetColor{Black}
\ArrowLine(130,0)(100,25){}{}\Vertex(100,25){2}
\SetColor{Black}
\ArrowLine(100,25)(130,55){}{}

\Photon(200,25)(250,25){5}{4}
\SetColor{Black}
\ArrowLine(170,0)(200,25){}{}
\SetColor{Black}
\ArrowLine(200,25)(170,55){}{}\Vertex(200,25){2}
\SetColor{Black}
\ArrowLine(300,0)(250,25){}{}
\SetColor{Black}
\ArrowLine(250,25)(300,55){}{}
\Photon(285,8)(285,45){4}{4}

\vspace*{0.8 true in}
\Text(75,-5)[]{(a)}
\Text(75,40)[]{$Z$}
\Text(35,50)[]{$b$}
\Text(35,05)[]{$s$}
\Text(120,55)[]{$s$}
\Text(120,-02)[]{$b$}

\Text(230,-5)[]{(b)}
\Text(225,40)[]{$Z$}

\Text(185,50)[]{$b$}
\Text(185,05)[]{$s$}

\Text(295,60)[]{$s$}
\Text(295,-10)[]{$b$}
\Text(300,25)[]{$W$}

\Text(265,45)[]{$u,c,t$}
\Text(265,5)[]{$u,c,t$}

\end{picture}
\end{center}
\caption{Feynman diagrams for $B_s -\bar B_s$ mixing in the model
with an extra vector like down quark, where the blob represents the
tree level flavor changing vertex.}
\end{figure}
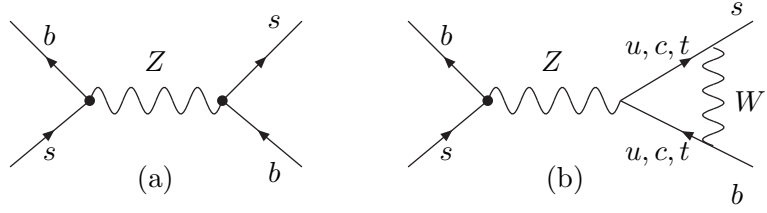


In the model with an extra vector like down quark there will be two
additional contributions to the $B_s - \bar B_s$ mixing amplitude.
The first one is induced by tree level FCNC mediated $Z$ boson, with
two non-standard (flavor-changing) $Z-b-s$ coupling as shown in
Figure-1(a) and the second contribution contains one non-standard
$Z-b-s$ coupling and one SM loop-induced $Z-b-s$ coupling as
depicted in Figure-1(b). With these new contributions the mass
difference between $B_s^H$ and $B_s^L$ deviates significantly from
its SM value.

To evaluate these two additional contributions, one can write from
Eq. (\ref{fcnc}) the effective FCNC mediated Lagrangian for $Zbs$
interaction as
 \be {\cal
L}_{FCNC}^Z=-\frac{g}{2 \cos \theta_W}U_{sb} \bar s_L \gamma^\mu b_L
Z_\mu\;.\label{ubs} \ee This gives the effective Hamiltonian induced by tree
level FCNC mediated $Z$ boson (Fig-1(a)) as \be {\cal H}_{eff}^Z=
\frac{G_F}{\sqrt 2} ~ U_{sb}^2 ~\eta_Z (\bar s_L \gamma^\mu b_L)
(\bar s_L \gamma_\mu b_L), \ee where $\eta_Z=(\alpha_s(m_Z))^{6/23}$
 is the QCD correction
factor. Using the matrix elements as defined in Eq. (\ref{vac}) we
obtain \be M_{12}^Z= \frac{G_F}{3 \sqrt 2} ~U_{sb}^2~\eta_Z B_s
f_{B_s}^2 M_{B_s}\;.\label{z} \ee

The effective Hamiltonian induced by the SM penguin at one vertex
and Z mediated FCNC coupling on the other (Figure-1(b)) is given as
\be {\cal H}_{eff}^{SM+Z}=\frac{G_F^2}{4 \pi^2}~
\lambda_t~\eta_{Zt}~ M_W^2 U_{sb} C_0(x_t) (\bar s b)_{V-A} (\bar s
b)_{V-A} \ee where $\eta_{Zt}$ is the QCD correction factor and  \be
C_0(x_t)=\frac{x_t}{8}\left ( \frac{x_t-6}{x_t-1}+\frac{3 x_t+2}{
(x_t-1)^2} \log x_t \right ). \ee This gives \be
M_{12}^{SM+Z}=\frac{G_F^2}{3 \pi^2}~ \lambda_t U_{sb}~\eta_{Zt}~
M_W^2 C_0(x_t) B_s f_{B_s}^2  M_{B_s}\;.\label{smz}\ee Thus, the
mass difference $\Delta M_s$ in this model can be given as \be\Delta
M_s=2\left | M_{12}^{SM}+M_{12}^Z+M_{12}^{SM+Z}\right |= \Delta
M_s^{\rm SM}\left |1+a \left (\frac{U_{sb}}{\lambda_t}\right )+ b
\left (\frac{U_{sb}}{\lambda_t} \right )^2 \right | \ee with \be a=4
\frac{C_0(x_t)}{ S_0(x_t)}, ~~~~~b=\frac{2 \sqrt 2 \pi^2}{ G_F M_W^2
S_0(x_t)}, \label{ab}\ee where we have assumed $\eta_t \approx
\eta_Z \approx \eta_{Zt}$. The coupling $U_{sb}$ characterizing the
$Z-b-s$
 strength is in general complex and can be parameterized as
 $U_{sb}=|U_{sb}| e^{i \phi_s}$, where $\phi_s$ is the new weak
 phase. The constraints on these parameters can be obtained
 using the recent measurement on $\Delta M_s$.

Since $V_{tb}V_{ts}^*=-|V_{tb}V_{ts}|e^{i \beta_s}$, we parametrize
\be
\frac{U_{sb}}{V_{tb}V_{ts}^*}=-\left 
|\frac{U_{sb}}{V_{tb}V_{ts}}\right | e^{i(\phi_s-\beta_s)}
\equiv -x~ e^{i(\phi_s-\beta_s)}\;.\label{x}
\ee
For numerical evaluation we use the CKM elements as
$|V_{tb}|=0.999176_{-0.000044}^{+0.000031}$,
$|V_{ts}|=0.03972_{-0.00077}^{+0.00115}$ \cite{ckmfitter},
$\beta_s =-1.1^\circ$,
the masses of $W$ boson and $t$ quark as $M_W=80.4$ GeV, $m_t=168$ GeV.
For $\Delta M_s$, we use the CDF result \cite{cdf1} $\Delta M_s=17.77 \pm
0.12~{\rm ps}^{-1} $ and for $\Delta M_s^{\rm SM}=19.30 \pm
6.68~{\rm ps}^{-1} $ \cite{lenz}, which yields $\Delta M_s/\Delta
M_s^{\rm SM}=0.92 \pm 0.32$.  Varying $(\Delta M_s/\Delta M_s^{\rm
SM})$ within its $1-\sigma $ range the allowed parameter space in
 the $\phi_s - |U_{sb}| $ plane is shown in Figure-2. From the figure
 it can be seen that for higher value of $|U_{sb}|$ the phase
 $\phi_s$ is very tightly constrained. However, for
 $|U_{sb}| \leq 0.0015$ there
is no constraint on the new weak phase $\phi_s$ i.e., the whole
range $0-2 \pi$ is allowed. 
The constraint on $|U_{sb}|$ 
obtained from $B \to X_s l^+ l^-$, i.e., $|U_{sb}|\leq 0.002$, 
\cite{bbm} is consistent with the constraint obtained
from $B_s - \bar B_s$ mixing.
We now use the allowed values of
$|U_{sb}|$ (i.e., we use  $|U_{sb}|\leq 0.002$ so that
constraints coming from both the observables will be satisfied)  
and $\phi_s$   to study  some anomalies associated with
$b\to s $ transitions. In particular, we would like to see whether
the constraints obtained above in the extended isosinglet down quark
model, consistent with $B_s-\bar B_s$ mixing, can also explain the
discrepancies in the modes $B_s\to \psi \phi$, $B_s \to \mu^+ \mu^-$,
$B\to \pi K$ and $B\to \phi K_s$.

\begin{figure}[htb]
\centerline{\epsfysize 2.5 truein \epsfbox{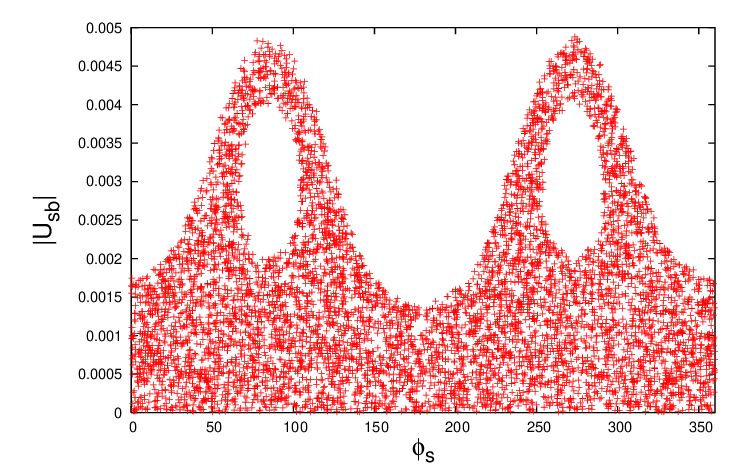}} \caption{The
$1-\sigma $ allowed range of  $(\Delta M_s/\Delta M_s^{\rm SM})$ in
the $\phi_s-|U_{sb}|$ plane.}
\end{figure}

\section{Mixing induced CP asymmetry in $B_s \to J/\psi
\phi~(S_{\psi \phi})$}

We now consider the effect of the isosinglet down quark on the
mixing induced CP asymmetry in $B_s \to J/\psi \phi$ mode. Recently
a very largish CP asymmetry has been measured by the CDF
collaboration \cite{cdf2} in the tagged analysis of $B_s \to J/\psi \phi$ with
value $S_{\psi\phi}$ $\in$ [0.23,~0.97].

Within the SM this asymmetry is expected to be vanishingly small,
which comes basically from $B_s -\bar B_s$ mixing phase. Since this
mode receives dominant contribution from $b \to c \bar c s$ tree
level transition, the NP contribution to its decay amplitude is
naively expected to be negligible. Therefore, the observed large CP
asymmetry is believed to be originating from the new CP violating
phase in $B_s -\bar B_s$ mixing.

Now parameterizing the new physics contribution to the $B_s -\bar
B_s$ mixing amplitude as \be
M_{12}=M_{12}^{SM}+M_{12}^Z+M_{12}^{SM+Z}=M_{12}^{\rm SM}~ C_{B_s}~
e^{2 i \theta_s}\;, \label{theta}\ee one can obtain
\be S_{\psi \phi}=-\eta_{\psi
\phi} \sin (2 \beta_s + 2 \theta_s)\;, \ee where $\beta_s$ is the
phase of $V_{ts}=-|V_{ts}|e^{-i \beta_s}$ and $\eta_{\psi \phi}$ is
the CP parity of the $\psi \phi$ final state. Taking $\eta_{\psi
\phi}=+1$ and $\beta_s\approx -1.1^\circ$ we obtain the  mixing induced CP
asymmetry as \be S_{\psi \phi}=\sin(2|\beta_s|-2 \theta_s).\label{spsi}
 \ee

\begin{figure}[htb]
\centerline{\epsfysize 2.5 truein \epsfbox{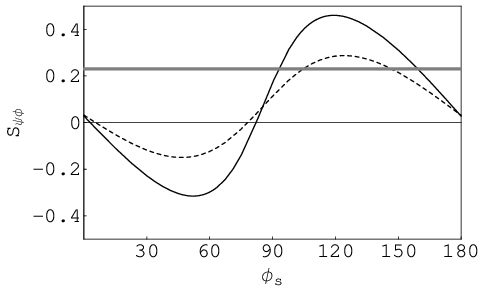}} \caption{
Variation of $S_{\psi \phi}$ with the new weak phase $\phi_s$ where
the solid and dotted lines are for $|U_{sb}|=0.002$ and 0.0015
respectively. The horizontal line represents the lower limit of the
experimental value. }
\end{figure}
Now substituting  the expressions for $M_{12}^{\rm SM}$,
$M_{12}^Z$ and $M_{12}^{SM+Z}$ from Eqs. (\ref{sm}),
(\ref{z}) and (\ref{smz}) in Eq. (\ref{theta}), we obtain
 the new CP-odd phase of $B_s - \bar B_s$ mixing as
 \be
 2\theta_s= \arctan\left (\frac{-a ~x \sin(\phi_s+|\beta_s|)
 +b~ x^2 \sin(2 \phi_s+2|\beta_s|)}{1 -a~ x \cos(\phi_s+|\beta_s|)
 +b~ x^2 \cos(2 \phi_s+2|\beta_s|)}  \right),\ee
 where $a$, $b$ and $x$ are defined in Eqs.
 (\ref{ab}) and (\ref{x}) respectively. 
In Figure-3 we show the variation of $S_{\psi \phi}$
(\ref{spsi}) with the
 new weak phase $\phi_s$ for two representative values of
 $|U_{sb}|$. From the figure it can be seen that the observed
 largish $S_{\psi \phi}$ can be explained in the model with an extra
 vector-like down quark for $|U_{sb}|\geq 0.0015$.

\section{$B_s \to \mu^+ \mu^-$}

Now let us consider the FCNC mediated leptonic transition  $B_s \to
\mu^+ \mu^-$. This decay  mode has attracted a lot of attention
recently since it is very sensitive to the structure of the SM and
potential source of  new physics beyond the SM. Furthermore, this
process is very clean and the only nonperturbative quantity involved
is the decay constant of $B_s$ meson which can be reliably
calculated by the well known non-perturbative methods such as QCD
sum rules, lattice gauge theory etc. Therefore, it provides a good
hunting ground to look for for new physics. The recent updated
branching ratio ${\rm Br}(B_s \to \mu^+ \mu^-)= (3.35 \pm 0.32)
\times 10^{-9}$ in the SM \cite{np} is well below the present
experimental upper limit \cite{hfag} \be  Br(B_s \to \mu^+
\mu^-) < 4.7 \times 10^{-8}\,. \ee This decay has been analyzed in
many beyond the SM scenarios in a number of papers \cite{bsmu}. Let
us start by recalling the result for $B_s \to \mu^+ \mu^-$ in
 standard model. The effective Hamiltonian describing
this process is \bea \label{one} {\cal{H}}_{eff} &=& \frac{G_F}{\sqrt 2}
\frac{\alpha }{\pi} V_{tb} V_{ts}^* \Bigg[ C_9 ~({\bar s}~
\gamma_\mu~ P_L ~b) ({\bar \mu}~ \gamma^\mu ~\mu) + C_{10}~({\bar
s}~ \gamma_\mu~ P_L ~b)({\bar \mu} ~\gamma^\mu ~\gamma_5 ~\mu)
\nonumber \\
&&
~~~~~~~- \frac{2 C_7~ m_b}{q^2} ({\bar s}i \sigma_{\mu \nu}
q^\nu P_R ~b)
( {\bar \mu}~\gamma^\mu ~ \mu)
\Bigg]\;,\label{ham}
\eea
where
$P_{L,R} = \frac{1}{2}~(1 \mp \gamma_5)$ and $q$ is the momentum
transfer.
$C_i$'s are the Wilson coefficients evaluated at the $b$ quark mass
scale in NLL order with values \cite{beneke}
\be
C_7=-0.308\;,~~C_9=4.154\;,~~C_{10}=-4.261\;.\label{wil}
\ee

To evaluate the transition amplitude one can generally adopt the vacuum
insertion method, where the form factors of the various currents
are defined as follows
\be
\langle 0~|~ {\bar s} ~\gamma^\mu~ \gamma_5 ~b~| B^0_s \rangle
= i f_{B_s} p^\mu_B\;,~~~\langle 0~| {\bar s} ~\gamma_5~ b| B^0_s \rangle
=  i f_{B_s} m_{B_s}\;,~~~
\langle 0 |~ {\bar s} ~\sigma^{\mu \nu} ~P_R ~b~ |B^0_s\rangle = 0\;.
\label{four}
\ee
Since $ p^\mu_B = p^\mu_+ + p^\mu_-$, the contribution from
$C_9$ term in Eq. (\ref {one})
will vanish upon contraction with the lepton bilinear, $C_7$
will also give zero
by (\ref{four}) and the remaining $C_{10}$ term will get a factor of
$2m_\mu$.
\par Thus the transition amplitude for the process is given as
\bea
\label{four1}
{\cal M}(B_s \to \mu^+ \mu^-)
&=& i\frac{G_F}{\sqrt 2}\frac{\alpha }{\pi} ~V_{tb} V_{ts}^*~
f_{B_s}~ C_{10}~ m_{\mu}
~ (\bar{\mu} \gamma_5 \mu)\;,
\eea
and the corresponding branching ratio is given as
\be
 Br(B_s \to \mu^+ \mu^-)
= \frac{ G_F^2~ \tau_{B_s}}{16 \pi^3}~\alpha^2~
f_{B_s}^2~ m_{B_s}~ m_{\mu}^2
~|V_{tb}V^*_{ts}|^2 ~C_{10}^2 ~\sqrt{1-\frac{4 m_{\mu}^2}{m_{B_s}^2}}\;.
\label{five}
\ee
Helicity suppression is reflected by the presence of $m_{\mu}^2$ in
(\ref{five}) which gives   a very small
branching ratio of $(3.35 \pm 0.32)\times
10^{-9}$ for $\mu^+ \mu^-$ \cite{np}.

Now let us analyze the decay modes $B_s \to \mu^+ \mu^-$ in the model with the
$Z$ mediated FCNC
occurring at the tree level.  The effective Hamiltonian for
 $B_s \to \mu^+ \mu^-$ is given as
\be
{\cal H}_{eff}= \frac{G_F}{\sqrt 2}~ U_{sb}~ [\bar s \gamma^\mu
(1-\gamma_5)b]
\left [\bar \mu( C_V^\mu \gamma_\mu  -C_A^\mu \gamma_\mu \gamma_5 ) \mu
\right ]\;,\label{ham1}
\ee
where $C_V^\mu $ and $C_A^\mu $  are the vector and axial vector
$Z \mu^+ \mu^-$
couplings, which are given as
\be
C_V^\mu= -\frac{1}{2}+2\sin^2 \theta_W\;,~~~~~~~
C_A^\mu= -\frac{1}{2}\;. \label{ca}
\ee
Since, the structure of the effective Hamiltonian (\ref{ham1}) in
this model is the same form  as that of the SM, like $\sim(V-A)(V-A)$ form,
therefore its
 effect on the various decay observables can be encoded by replacing the
SM Wilson coefficients $C_9$ and $C_{10}$ by
\bea C_9^{eff} = C_9+\frac{2 \pi}{\alpha}\frac{ U_{sb}
C_V^\mu}{ V_{tb}V_{ts}^*}\;,~~~~ C_{10}^{eff} = C_{10}-\frac{2
\pi}{\alpha}\frac{ U_{sb}C_A^\mu}{ V_{tb}V_{ts}^*}\;.\label{nph} \eea

Thus, one can obtain the branching ratio including the NP
contributions by substituting $C_{10}^{eff}$ from (\ref{nph}) in
(\ref{five}). Now varying the the value $|U_{sb}|$ between 0 and
$1.5 \times 10^{-3}$ and the phase $\phi_s$ between $(0-360)^\circ$
the branching ratio for $B_s \to \mu^+ \mu^-$ is shown in Figure-4.
From the figure one can conclude that the branching ratio of $B_s
\to \mu^+ \mu^-$ in this model can be significantly enhanced
from its SM value. Observation of this mode  in the upcoming
experiments will provide additional constraints on the new physics
parameters.

\begin{figure}[htb]
\centerline{\epsfysize 2.5 truein \epsfbox{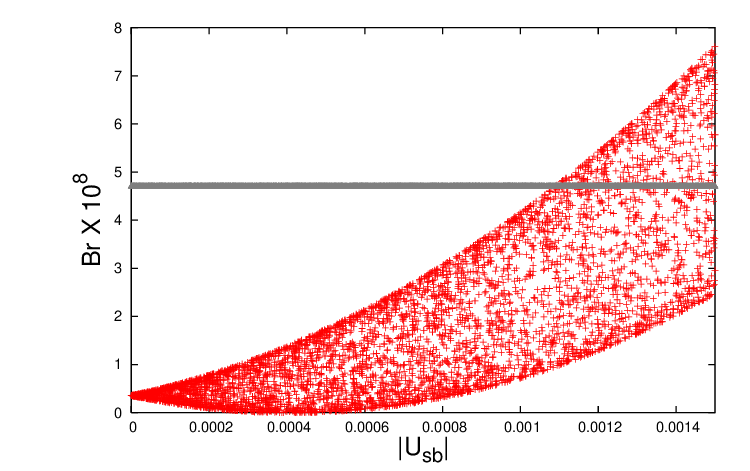}} \caption{The
allowed range of the branching ratio for $B_s \to \mu^+ \mu^-$
process in the $Br - |U_{sb}|$ plane. The horizontal line represents
the experimental upper limit.}
\end{figure}

\section{$\Delta A_{CP}(K \pi) $ Puzzle}

The $\Delta A_{CP}(K \pi)$ puzzle refers to the difference in direct
CP asymmetries in $B^- \to
\pi^0 K^-$ and $\bar B^0 \to \pi^+ K^-$ modes. These two modes
receive similar dominating contributions from tree and QCD penguin
diagrams and hence one would naively expect that these two channels
will have the same direct CP asymmetries i.e., ${\cal A}_{\pi^0
K^-}= {\cal A}_{\pi^+ K^-}$. In the QCD factorization approach, the
difference between these asymmetries  is found to be \cite{soni}
 \be \Delta A_{CP} ={\cal A}_{ K^- \pi^0} -{\cal A}_
{ K^- \pi^+} =(2.5 \pm
 1.5)\%
 \ee
 whereas the corresponding experimental value is \cite{hfag}
\be \Delta A_{CP} =(14.8 \pm
 2.8)\%\;,
 \ee
which yields nearly  $4 \sigma$ deviation.

In the SM, the relevant
effective Hamiltonian describing the decay modes $B^- \to \pi^0 K^-
$ and $\bar B^0 \to \pi^+ K^-$  is given by
\begin{equation}
{\cal H}_{eff}^{SM} = \frac{G_F}{\sqrt{2}}\left[  V_{ub}
V_{us}^*(C_1O_1+C_2 O_2)- V_{tb}V_{ts}^*\sum_{i=3}^{10} C_i O_i
\right]\;,
\end{equation}
where $C_i$'s are the Wilson coefficients evaluated at the $b$ quark mass
scale and $O_i$'s are the four-quark current operators.

Thus, one can obtain the transition amplitudes in the QCD
factorization approach as \cite{qcdf},
where the CKM unitarity $\lambda_u + \lambda_c +
\lambda_t=0$ has been used
 \bea \sqrt 2 A(B^- \to
\pi^0 K^-)&=&\lambda_u \Big(A_{\pi \bar K}(\alpha_1+\beta_2)+A_{\bar
K \pi}\alpha_2 \Big)\nn\\
&+&\sum_{q=u,~c}\lambda_q\Big(A_{\pi \bar
K}(\alpha_4^q+\alpha_{4,EW}^q+\beta_3^q+\beta_{3,EW}^q)+\frac{3}{2}A_{\bar
K \pi}\alpha_{3,EW}^q \Big) \eea and \bea A(\bar B^0 \to  \pi^+
K^-)=\lambda_u \Big(A_{\pi \bar K}~\alpha_1 \Big) +
\sum_{q=u,~c}\lambda_q A_{\pi
\bar
K}\Big(\alpha_4^q+\alpha_{4,EW}^q+\beta_3^q-\frac{1}{2}
\beta_{3,EW}^q \Big),
\eea where \be  A_{\pi \bar K}= i\frac{G_F}{\sqrt 2}M_B^2 F_0^{B\to
\pi}(0)f_K~~~~~{\rm and}~~~~A_{ \bar K \pi}= i\frac{G_F}{\sqrt 2}M_B^2
F_0^{B\to K}(0)f_{\pi}\;. \ee
The parameters $\alpha_i$'s and $\beta_i$'s are related to the 
Wilson coefficients $C_i$'s and the corresponding expressions can be 
found in \cite{qcdf}.

To account for this discrepancy here we consider the effect
of the extra isosinglet down quark. As discussed earlier,
in this model the $Z$ mediated FCNC interaction is
introduced at the tree level as shown in Eq. (\ref{ubs}).
Because of the new interactions the effective Hamiltonian describing
$b \to s \bar s s$ process receives the additional
contribution given  as \cite{desh}, \be {\cal
H}_{eff}^Z= - \frac{G_F}{\sqrt 2}[ \tilde C_3 O_3 +
\tilde C_7 O_7 + \tilde C_9 O_9]\;, \ee where the four-quark
operators $O_3$, $O_7$ and $O_9$ have the same structure as the SM
QCD and electroweak penguin operators and the new Wilson
coefficients $\tilde C_i$'s at the $M_Z$ scale are given by
\bea \tilde C_3(M_Z) &=& \frac{1}{6} U_{sb},\nn\\
\tilde C_7(M_Z) &=& \frac{2}{3} U_{sb}
\sin^2 \theta_W,\nn\\
\tilde C_9(M_Z) &=& -\frac{2}{3} U_{sb}
(1- \sin^2 \theta_W).
 \eea
These new Wilson coefficients will be evolved from the $M_Z$ scale
to the $m_b$ scale using renormalization group equation given in
\cite{wilson}, as \be \vec C_i(m_b)=U_5(m_b, M_W, \alpha ) \vec
C(M_W)\;, \ee where $\vec C$ is the $10 \times 1$ column vector of
the Wilson coefficients and $U_5$ is the five flavor $10 \times 10$
evolution matrix. The explicit forms of $\vec C(M_W)$ and $U_5(m_b,
M_W, \alpha)$ are given in \cite{wilson}.as described earlier.
Because of the RG evolution these three Wilson coefficients generate
new set of Wilson coefficients $\tilde C_i (i=3,\cdots, 10)$ at the
low energy regime (i.e., at the $m_b$ scale) as presented in
Table-1, where we have used $\sin^2 \theta_W=0.231$.

\begin{table}[htbp]
\begin{center}
\begin{tabular}{|c|c|c|c|c|c|c|c|}
\hline
$\tilde C_3$ & $\tilde C_4$ & $\tilde C_5$&$\tilde C_6$&
$\tilde C_7$&$\tilde C_8$&$\tilde C_9$&$\tilde C_{10}$ \\
\hline 0.19$U_{sb} $ & $-0.066U_{sb}$ &0.009$U_{sb}
$& $-0.031U_{sb} $&0.145$U_{sb}$ &0.053$U_{sb}$ & $-0.566U_{sb}$
&0.127$U_{sb}$\\
\hline
\end{tabular}
\caption{Values of the new Wilson coefficients at the $m_b$ scale.
 } \label{tab2}
\end{center}
\end{table}
As discussed earlier, due to the presence of the additional isosinglet
down quark the unitarity condition becomes $\lambda_u + \lambda_c 
+\lambda_t = U_{sb}$.  Thus, replacing $\lambda_t =U_{sb}
-(\lambda_u + \lambda_c)$,  one can write
the transition amplitudes including the new contributions as
 \bea \sqrt2 A(B^- \to
\pi^0 K^-)&=&\lambda_u \Big(A_{\pi \bar K}(\alpha_1+\beta_2)+A_{\bar
K \pi}\alpha_2 \Big)\nn\\
&+&\sum_{q=u,~c}\lambda_q\Big(A_{\pi \bar
K}(\alpha_4^q+\alpha_{4,EW}^q+\beta_3^q+\beta_{3,EW}^q)+\frac{3}{2}A_{\bar
K \pi}\alpha_{3,EW}^q \Big)\nn\\
&-&U_{sb}\Big(A_{\pi \bar K}(\Delta \alpha_4+\Delta
\alpha_{4,EW}+\Delta \beta_3+\Delta \beta_{3,EW})+\frac{3}{2}A_{\bar
K \pi}\Delta\alpha_{3,EW} \Big) \eea and \bea A(\bar B^0 \to  \pi^+
K^-)&=&\lambda_u \Big(A_{\pi \bar K}~\alpha_1 \Big) 
+\sum_{q=u,~c}\lambda_q A_{\pi
\bar K}\Big(\alpha_4^q+\alpha_{4,EW}^q+\beta_3^q-\frac{1}{2}
\beta_{3,EW}^q \Big)\nn\\
&-& U_{sb}~ A_{\pi \bar K}\Big(\Delta \alpha_4+\Delta
\alpha_{4,EW}+\Delta \beta_3-\frac{1}{2}\Delta \beta_{3,EW} \Big),
\eea where
$\Delta \alpha_i$'s and $\Delta \beta_i$'s are related to the 
modified  Wilson 
coefficients $\Delta C_i = \tilde
C_i(m_b)+ C_i^t(m_b)$, where $C_i^t(m_b)$'s are the values of the 
Wilson coefficients at the $m_b$ scale due to $t$ quark exchange.

Thus, including the new contributions one can symbolically represent
these amplitudes as \bea Amp= \lambda_u A_u +\lambda_c A_c
-U_{sb}~A_{new} .\eea $\lambda$'s and $U_{bs}$ contain the weak phase
information and $A_i$'s are associated with the strong phases. Thus
one can explicitly separate the strong and weak phases and write the
amplitudes as \bea Amp=\lambda_c A_c\Big[1+ r~ a~
e^{i(\delta_1-\gamma)}-r'~ b~ e^{i(\delta_2+\phi_s)}], \eea where
$a=|\lambda_u/\lambda_c|$, $b=|U_{sb}/\lambda_c|$, $-\gamma$ is the
weak phase of $V_{ub}$ and $\phi_s$ is the weak phase of $U_{sb}$.
$r=|A_u/A_c|$, $r'=|A_{new}/A_c|$, and $\delta_1$ ($\delta_2$) is
the relative strong phases between $A_u$ and $A_c$ ($A_{new}$ and
$A_c$) Thus from the above amplitudes one can obtain the direct CP
asymmetry parameter as \bea A_{CP}=\frac{2\Big[ra \sin \delta_1 \sin
\gamma +r' b \sin \delta_2 \sin \phi_s + r r' a b
\sin(\delta_2-\delta_1) \sin
(\gamma+\phi_s)\Big]}{\Big[{\cal{R}}+2(ra \cos \delta_1 \cos
\gamma-2r' b \cos \phi_s \cos \delta_2 - 2 r r' a b
\cos(\gamma+\phi_s) \cos(\delta_2-\delta_1))\Big]} \eea where
${\cal{R}}=1+(ra)^2+(r'b)^2$.

For numerical evaluation,  we use input parameters as given in the
S4 scenario of QCD factorization approach. For the CKM matrix
elements we use the values from \cite{pdg}, extracted from
direct measurements and 
$\gamma=\left (67_{-25}^{+32} \right )^\circ$ \cite{ckmfitter}.
The particle masses are taken from \cite{pdg}.
We vary the $|U_{bs}|$ in the range $0 \leq
|U_{sb}| \leq 0.002$ and the corresponding phase between $30^\circ
\leq \phi_s \leq 150^\circ$ and the allowed region in $\Delta
A_{CP}$ and $|U_{sb}|$ plane is shown in the Fig.-5.
From the figure it can be seen that the observed
 $\Delta A_{CP}$ can
be accommodated in the VLDQ model.

\begin{figure}[htb]
\centerline{\epsfysize 2.5 truein \epsfbox{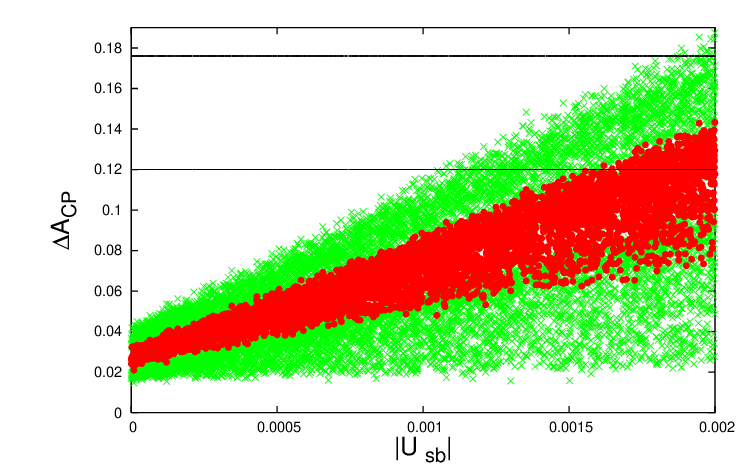}} \caption{The
allowed range  of the CP asymmetry difference ($\Delta A_{CP}$) in
the ($\Delta A_{CP}-|U_{sb}|$) plane as shown by the red region. The
30 \% error bars are due to hadronic uncertainties and shown by
green bands. The horizontal lines correspond to the experimentally
allowed $1-\sigma$ range. }
\end{figure}

\section{$S_{\phi K_s}$}

Next we consider the decay mode $\bar B^0 \to \phi K^0$. In the SM,
it proceeds through the quark level transition $b \to s \bar s s$
and hence the mixing induced CP asymmetry in this mode ($S_{\phi
K}$) is expected to give the same value as that of the  $B \to
J/\psi K_s$ with an uncertainty of around $5 \%$. However, the
present world average of this parameter is $S_{\phi K}=
0.44_{-0.18}^{+0.17}$  \cite{hfag}, which has nearly 
2.4$\sigma$ deviation from the 
corresponding
$S_{\psi K_s}$, with $S_{\phi K_s} < S_{\psi K_s}$. We would like to
see whether the
 model with an extra vector like down quark can 
account for this discrepancy.

In this  model one can write the amplitude for this process,
analogous to $B \to \pi K$ processes, as \bea A(\bar B^0 \to \bar
K^0 \phi)&=&A_{\bar K \phi} \Big[\sum_{q=u,c} \lambda_q\left
(\alpha_3^q+\alpha_4^q+\beta_3^q-\frac{1}{2}
\left (\alpha_{3, EW}^q+\alpha_{4, EW}^q+\beta_{3, EW}^q \right)\right )\nn\\
&-&U_{sb}\left (\Delta\alpha_3+\Delta\alpha_4+ \Delta
\beta_3-\frac{1}{2} \left (\Delta \alpha_{3, EW}+\Delta \alpha_{4,
EW}+\Delta \beta_{3, EW}\right )\right )\Big], \eea 
with $A_{\bar K^0 \phi}=-2 m_\phi (\epsilon_\phi \cdot p_B) F_+^{B \to K}
(0) f_\phi$,
which again can
be expressed as \bea A(\bar B^0 \to \bar K^0 \phi)=\lambda_u
A_u^\prime+\lambda_c A_c^\prime -U_{sb}~A_{new}^\prime=\lambda_c
A_c^{\prime}[1+r_1 e^{i(\delta-\gamma)}-r_1^\prime b ~e^{i
\phi_s}e^{i \delta'}], \eea where \be r_1=|A_{u}^\prime/A_c^\prime|,
~~~~\delta=Arg(A_{u}^\prime/A_c^\prime)~~~r_1^\prime
=|A_{new}^\prime/A_c^\prime|,
~~~~\delta'=Arg(A_{new}^\prime/A_c^\prime)\;. \ee Thus one can obtain
the expression for mixing induced CP asymmetry parameter as \be
S_{\phi K}= \frac{X}{{\cal R}'+2r_1 a \cos \delta \cos
\gamma-2 r_1' b \cos \delta' \cos \phi_s-2 r_1 r_1' a b
\cos(\delta-\delta') \cos(\gamma+\phi_s)}\;, \ee where
${\cal R}'=1+(r_1 a)^2+(r_1' b)^2$ and \bea X &=&
\sin 2 \beta+ 2 r_1 a \cos \delta \sin(2 \beta+\gamma)- 2 r_1' b
\cos \delta' \sin(2 \beta- \phi_s) +(r_1 a)^2\sin(2 \beta +2 \gamma)
\nn\\ &+&(r_1' b)^2\sin(2 \beta -2 \phi_s)-2 r_1 r_1' a b
\cos(\delta - \delta') \sin(2 \beta+\gamma -\phi_s). \eea

For numerical evaluation we use the input parameters as given in S4
scenario of QCD factorization. Using the CKM elements, as discussed
earlier, alongwith $\beta=(21.1 \pm 0.9)^\circ$ \cite{hfag},
 the variation of $S_{\phi K}$ with $\phi_s$ for different
values of $|U_{sb}|$ is shown in Figure-6. From the figure it
can be seen that the experimental value of $S_{\phi K}$ can be
accommodated in this model.

\begin{figure}[htb]
\centerline{\epsfysize 2.5 truein \epsfbox{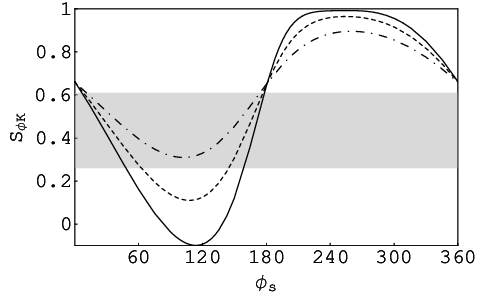}} \caption{The
variation of $S_{\phi K}$ (in S4 scenario) with the new weak phase
$\phi_s$, where the dot-dashed,  short-dashed and solid
 curves are for  $|U_{sb}| = 0.001,$ 0.0015 and 0.002.
The horizontal band corresponds to experimental allowed
 1$\sigma $ range.}
\end{figure}

\section{Summary and Conclusion}

Recent result of $B_s-\bar {B_s}$ mixing has created a lot of
attention in B decays and furthermore it is also claimed in the
literature that it could be the first evidence of physics beyond the
SM in the b-sector. Of course, there are many candidate beyond the
SM scenarios which can explain such a dicrepancy but here we will
employ the model with an extended isosinglet down quark to study the
same and explore whether other seemingly problematic deviations in
the $b\to s$ sector, as indicated by the data at present, can also
be explained simultaneously.

A minimal extension of the SM with only addition of an extra
isosinglet down quark in a vector like representation of the SM
gauge group that induces FCNC couplings in the Z boson couplings. These
models naturally arise for instance as the low energy limit of an
$E_6$ grand unified theory. From the phenomenlogical point of view
models with isosinglet quarks provide the simplest self-consistent
framework to study deviations of 3$\times$ 3 unitarity of the CKM
matrix as well as flavor changing neutral currents at the tree
level.

As stated earlier, we impose the extended isosinglet down quark
model to explain the deviation of $B_s-\bar {B_s}$ mixing from that
of the SM expectation and obtained the constraints on the parameters
of the new physics model and checked whether these severely
constrained parameters still can explain other $b\to s$ processes,
which appear to be not in agreement with the SM expectations.

Recently, CDF observed that the mixing induced parameter ($S_{\psi
\phi}$) for the decay mode $B_s\to \psi \phi$ appears to be not in
agreement with the SM expectation. In the SM, the value of  $B_s\to
\psi \phi$ is vanishingly small but the experiment has found a
rather large value which might be an indication of new physics. We
applied the constraints of the new physics model, obtained from the
$B_s-\bar {B_s}$ mixing, to see whether one can explain the same. It
can be seen from the figure-3 that one can explain the discrepany in
the NP model under consideration.

Next we consider the decay mode $B_s\to \mu^+\mu^-$, which is believed
to be a very clean mode and only the upper limit ($< 4.7\times$
$10^{-8}$) on its branching ratio has been obtained so far which is
much larger than the SM value. We used the contraints of the
isosinglet down quark model and see that (figure-4) a huge
enhancement can be possible due to its effect and can reach the
upper limit obtained by the experiment.

Thereafter, we considet the $\pi K$ puzzle, which is basically the
difference of direct CP asymmetry parameters, represented by $\Delta
A_{CP}(K\pi)$, of the modes $B^-\to \pi^0K^-$ and $\bar B^0 \to
\pi^+K^-$. In the SM  value of $\Delta A_{CP}(K\pi)$ is expected to
be close to zero whereas the experimental value is found
to be around 15\%. Invoking the new physics constraints,
obtained before, we have shown that the observed asymmetry 
can be obtained in this scenario. 

Finally, we consider the long standing problem of $S_{\phi K_s}$
corresponding to the decay mode $B\to \phi K_s$, which has about 2.5
sigma deviation from that of the $S_{\psi K_s}$. This large
deviation is belived to be the due to beyond the SM physics. We
employed the NP model under consideration and found that it can easily
explain such a disrepancy (figure-6).

To conclude, in this paper we employed the model with an extended
isosinglet down quark to constrain the parameters of the model using
the $B_s-\bar {B_s}$ mixing result. Thereafter, we checked 
whether deviations in other
$b\to s$ modes, namely, $B_s\to \psi \phi$, $B_s\to \mu^+\mu^-$, $B\to
\pi K$ and $B\to \phi K_s$ can also be understood in this
modeland found that the new physics parameters allowed by 
$B_s -\bar B_s$ mixing result can explain these discrepancies successfully.
 With more data in the future we will have a better understanding 
of these problems and possibly we shall be able to ascertain the 
nature of the new physics or else rule out some of
the existing beyond the SM scenarios, 
which appear to be allowed at present.

\acknowledgments The work of RM was partly supported by Department
of Science and Technology, Government of India, through grant Nos.
SR/S2/HEP-04/2005 and SR/S2/RFPS-03/2006. AG would like to thank
Council of Scientific and Industrial Research and Department of
Science and Technology, Government of India, for financial support.

\end{document}